\documentclass[aps,prd,twocolumn,showpacs,superscriptaddress,nofootbib,floatfix,longbibliography, preprintnumbers]{revtex4-1}

\usepackage[utf8]{inputenc}
\usepackage{amsmath,amssymb,amsfonts}
\usepackage{mathrsfs}
\usepackage{graphicx}
\usepackage[dvipsnames]{xcolor}
\usepackage{bm}
\usepackage{siunitx}
\usepackage{physics}
\usepackage{hyperref}
\usepackage{ulem}
\usepackage{orcidlink}
\usepackage[capitalise]{cleveref}
\usepackage{subcaption} 
\hypersetup{
    colorlinks=true,
    linkcolor=NavyBlue,
    citecolor=NavyBlue,
    urlcolor=NavyBlue
}

\begin{document}

\preprint{KEK-QUP-2026-0005, KEK-TH-2819} 

\title{Casimir Geometry as a Probe of Short Range Forces}

\author{Xiaolin Ma~\orcidlink{0009-0007-1994-9493}}  
\altaffiliation{\href{mailto:xlmphy@post.kek.jp}{xlmphy@post.kek.jp}, Corresponding} 
\affiliation{International Center for Quantum-field Measurement Systems for Studies of the Universe and Particles (QUP, WPI),
High Energy Accelerator Research Organization (KEK), Oho 1-1, Tsukuba, Ibaraki 305-0801, Japan}

\author{Volodymyr Takhistov}  
\altaffiliation{\href{mailto:vtakhist@post.kek.jp}{vtakhist@post.kek.jp}, Corresponding} 
\affiliation{International Center for Quantum-field Measurement Systems for Studies of the Universe and Particles (QUP, WPI),
High Energy Accelerator Research Organization (KEK), Oho 1-1, Tsukuba, Ibaraki 305-0801, Japan}
\affiliation{Theory Center, Institute of Particle and Nuclear Studies (IPNS), High Energy Accelerator Research Organization (KEK), \\ Tsukuba 305-0801, Japan
}
\affiliation{Graduate University for Advanced Studies (SOKENDAI), 
1-1 Oho, Tsukuba, Ibaraki 305-0801, Japan}
\affiliation{Kavli Institute for the Physics and Mathematics of the Universe (WPI), UTIAS, \\The University of Tokyo, Kashiwa, Chiba 277-8583, Japan}

\author{Hideo Iizuka}
\altaffiliation{\href{mailto:hiizuka@mosk.tytlabs.co.jp}{hiizuka@mosk.tytlabs.co.jp}}
\affiliation{International Center for Quantum-field Measurement Systems for Studies of the Universe and Particles (QUP, WPI),
High Energy Accelerator Research Organization (KEK), Oho 1-1, Tsukuba, Ibaraki 305-0801, Japan}

\date{\today}

\begin{abstract}
Casimir force searches provide among the most sensitive laboratory probes of   new short range interactions. Existing constraints rely almost exclusively on a single geometry. We show that Casimir geometry constitutes an independent observable, as  Yukawa-type  interactions and Casimir background exhibit  different geometric scaling for bulk forces and surface quantum effects.  We derive the first constraints from sphere-sphere and plate-plate geometries, thereby completing the  canonical set of Casimir geometries,   obtaining the most stringent Casimir-based bounds for $\lambda \lesssim 10^{-8}~\mathrm{m}$. Our results establish geometry as a new handle for systematic searches for short range forces.  
\end{abstract}

\maketitle

\textit{Introduction.--} 
Tests of possible new short range forces that manifest as deviations from Newton's inverse square law of gravity at short distances play an essential role in precision frontier searches for new fundamental physics~\cite{Safronova:2017xyt,Adelberger:2009zz,Murata:2014nra}. A broad range of theories beyond the Standard Model predict that deviations from Newtonian gravity can appear at submicron scales, including scenarios with extra spatial dimensions or new light bosons~\cite{Adelberger:2003zx,Arkani-Hamed:1998jmv,PhysRevD.59.086004,Damour:1994zq,Dimopoulos:1996gy,Antoniadis:1998ig}.  

While there is extensive experimental search program for novel short-range forces, including torsion-pendulum experiments~\cite{Talmadge:1988qz,Spero:1980zz,Hoskins:1985tn,Tan:2020vpf,Tan:2016vwu,Lee:2020zjt} as well as  particle scattering and collider experiments~\cite{Haddock:2017wav,Pokotilovski:2006up,Kamiya:2015eva,Kamiya:2020pyw,Nesvizhevsky:2007by,Voronin:2023muq,Salumbides:2013dua,Germann:2021koc,Lemos:2021xpi}~(see Ref.~\cite{Cong:2024qly} for review),  probing parameter space at distance separations below a few microns is notoriously challenging due to the presence of the Casimir effect. This quantum electrodynamic force between objects that typically dominates at these scales~\cite{PLUNIEN198687,PhysRevD.72.021301,Klimchitskaya:2009dw} has been long predicted~\cite{Casimir:1948dh} and definitively established experimentally under various conditions~\cite{Lamoreaux:1996wh}.
Precision Casimir force experiments~\cite{Chen:2014oda,Decca:2007yb,Mohideen:1998iz,Sushkov:2011md,Geraci:2008hb,Klimchitskaya:1999gg,Harris:2000zz} thus provide particularly powerful laboratory probes of new fundamental physics at such scales.

\begin{figure*}[t]
\begin{center}
    \begin{subfigure}{0.3\linewidth}
    \centering
    \includegraphics[width=\linewidth]{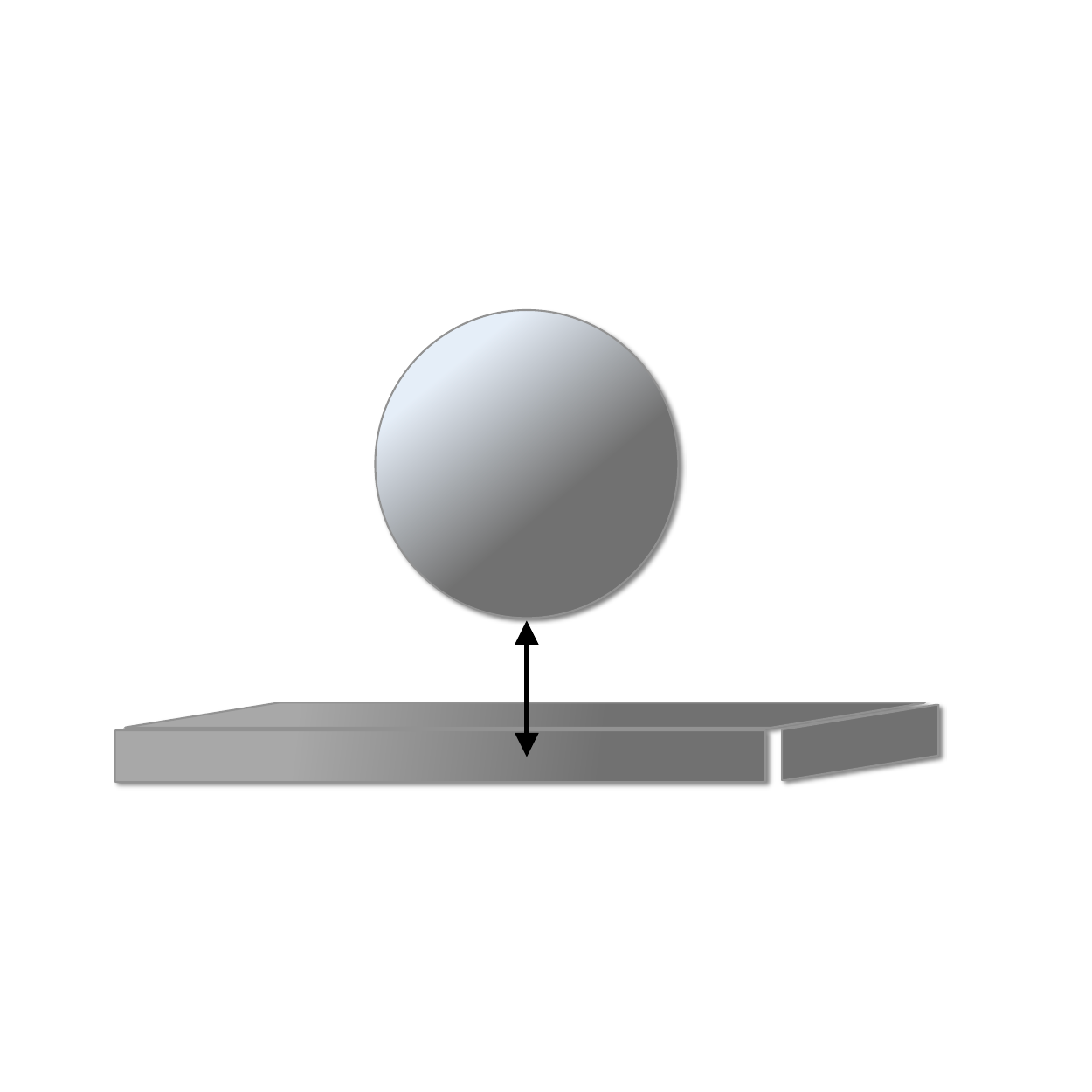}
    \caption{\textbf{Sphere-plate (s-p)} ~\newline $F'_{\rm Yuk} \sim \lambda e^{-d/\lambda} ~( +~{\rm corr.})$}
\end{subfigure}
\hfill
\begin{subfigure}{0.3\linewidth}
    \centering
    \includegraphics[width=\linewidth]{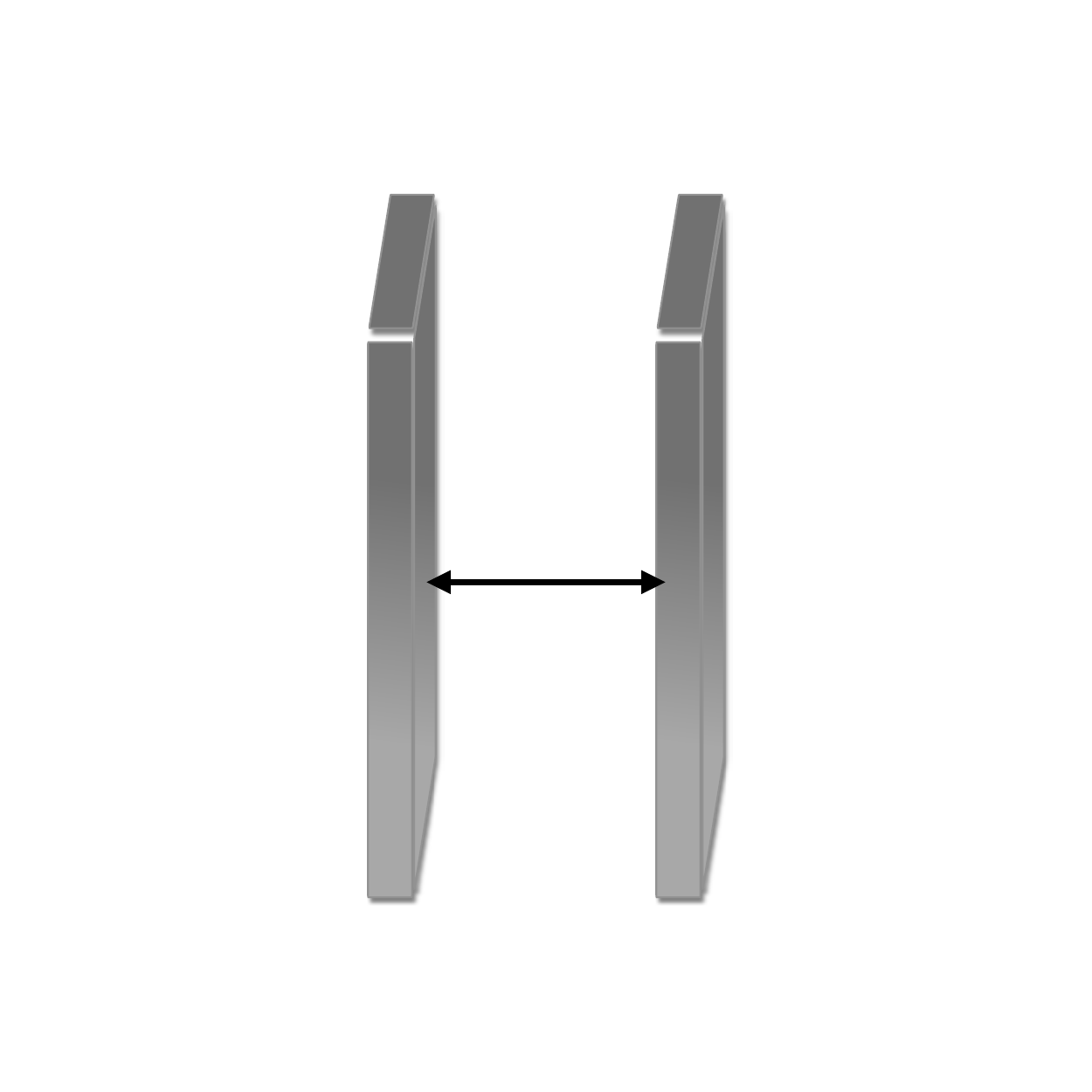} 
    \caption{\textbf{Plate-plate (p-p)} ~\newline    $F'_{\rm Yuk} \sim \lambda e^{-d/\lambda}$} 
\end{subfigure}
\hfill
\begin{subfigure}{0.3\linewidth}
    \centering
    \includegraphics[width=\linewidth]{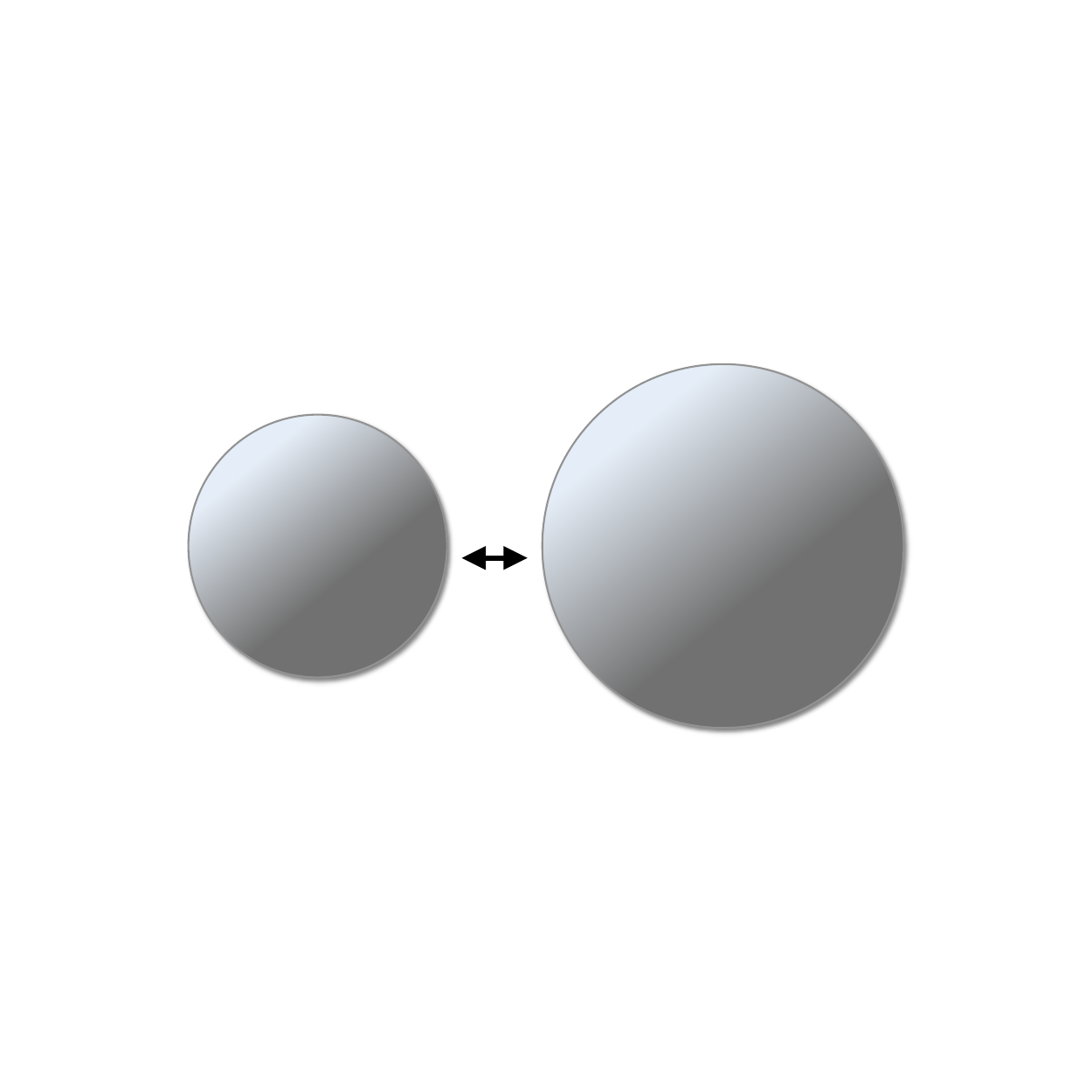}
    \caption{\textbf{Sphere-sphere (s-s)} ~\newline     $F'_{\rm Yuk} \sim \lambda^2 e^{-d/\lambda} ~( +~{\rm corr.})$}
\end{subfigure}
\end{center}
    \caption{Schematic illustrations of the Casimir force experimental configurations for Yukawa-type new force search: (a) sphere-plate (s-p) geometry, (b)  plate-plate (p-p) geometry and (c) sphere-sphere (s-s) geometry. Scaling of Yukawa-type force gradient for each geometry is shown, with ``corr.'' denoting correction effects. The distinct $\lambda$-scaling and corrections across geometries underlies the sensitivity to new forces demonstrated in this work.}
    \label{fig:schematic}
\end{figure*}

Nearly all Casimir searches for short range forces have relied on measurements between a sphere and a plate (s-p)~\cite{Chen:2014oda,Mohideen:1998iz,Bordag:2001qi}. Although other Casimir geometries have been realized experimentally, their potential as probes of novel interactions has remained largely overlooked. Since Yukawa-type forces depend on the bulk mass distribution  whereas the Casimir interactions are dominated by surface effects, different geometries can exhibit parametrically different sensitivities to hypothetical short range interactions. This implies that Casimir measurements performed in different geometries do not merely provide complementary sensitivity, but introduce an independent observable through their distinct scaling behavior, helping discrimination of hypothetical bulk forces from the Casimir background. Here we demonstrate that plate-plate (p-p) and sphere-sphere (s-s) geometries provide distinct and complementary probes of short range forces.

The p-p geometry is particularly challenging to analyze due to requirement of maintaining strict parallelism between macroscopic surfaces at sub-micron separations. For ideal conductors separated by distance $d$ the Casimir pressure scales as $P_C(d) \propto d^{-4}$, and thus even microradian tilts can dramatically affect the measured force~\cite{Garrett:2018zpo}. Nevertheless, this configuration is of particular significance as it corresponds directly to the original proposed setup by Casimir in 1948~\cite{Casimir:1948dh} and allows comparison with theoretical calculations without relying on geometric approximations such as the proximity force approximation (PFA) typically used for curved geometries. Further, p-p configurations hold potential advantage in new force searches as increasing the interacting surface area amplifies the total force above detector noise. This enables measurements at larger distance separations where the Casimir force effects are reduced and sensitivity to new longer range interactions is enhanced~\cite{Sedmik:2020cfj,Almasi:2015zpa}. Upcoming CANNEX experiment aims to implement the p-p geometry for tests of new forces with enhanced sensitivity~\cite{Sedmik:2021iaw}.

Conversely, the s-s geometry presents distinct experimental challenges, primarily the precise determination of the point of closest approach between the two curved surfaces~\cite{PhysRevLett.120.040401}. Any lateral displacement deviations can modify the effective separation and distort the inferred force. However, the high degree of symmetry of this configuration provides important advantages. The force interaction is localized near the nearest points of the spheres, which reduces sensitivity to large scale surface roughness~\cite{MaiaNeto:2005cxc} and spatial variations of electrostatic patch potentials~\cite{Kim:2009mr,Behunin:2011gj}. Furthermore, this significant force localization effectively eliminates the surface edge effects and macroscopic alignment ambiguities that can affect plate-based geometries.

We show that Casimir measurements performed in sphere-sphere and plate-plate geometries provide new probes of Yukawa-type short-range interactions beyond the conventional sphere-plate setup. By reanalyzing precision measurements in these complementary geometries, we derive the first constraints from each and demonstrate that the geometry enables an independent handle for separating hypothetical bulk forces from the dominant Casimir background. The absence of additional signals beyond the measured Casimir force allows us to derive exclusion limits on non-Newtonian interactions in the nanometer-micron range.

\textit{Deviations from Newtonian gravity.--}
A broad class of hypothetical short-range interactions   can be parametrized by a Yukawa-type modification of the Newtonian gravitational potential  between two
masses $m_1$ and $m_2$ as~\cite{Talmadge:1988qz,Adelberger:2003zx}
\begin{equation}
    V(r) = -G \frac{m_1 m_2}{r} \left( 1 + \alpha e^{-r/\lambda} \right),
    \label{eq:yukawa}
\end{equation}
where $G$ is the gravitational constant, $\alpha$ is the dimensionless coupling strength of the new interaction relative to gravity, and $\lambda$ is its characteristic range. Microscopically, $\lambda$ can be related to the mass of a new force mediating particle $\phi$ of mass $ m _{\phi} = 1/\lambda$~\cite{Fischbach:1996eq}.
Exchange of scalar or vector bosons between non-relativistic fermions with coupling strength $g_{S, V}$ can then be generically related to Eq.~\eqref{eq:yukawa} as $\displaystyle \pm(g_{\rm S,V}^2/4\pi)\leftrightarrow Gm_1m_2\alpha$, where $-$ and $+$ correspond to scalar and vector interactions respectively, depending on the source particle and the   geometry~\cite{Adelberger:2006dh}. 
We adopt natural units $\hbar = c=k_B=1$.

The experiments considered here, schematically illustrated in Fig.~\ref{fig:schematic}, measure the force  gradient  $F' = \partial F/\partial d$ with  distance $d$  rather than the static force $F$ directly. 
Accordingly, our analysis focuses on new physics effects resulting from gradient of the Yukawa-type interactions and leading to deviations from observable Casimir force.

We first consider the simpler p-p geometry as originally proposed for Casimir experiments. The Yukawa force between two large parallel plates of thicknesses  $t_1$ and $t_2$, mass densities $\rho_1$ and $\rho_2$, and overlapping area $S$, separated by a distance $d$ and assuming infinitely extended plates is approximately given by~\cite{Decca:2005qz,Ema:2023kvw}
\begin{align} \label{eq:ppapp}
F_{\rm Yuk-i,pp}(d) =&~ 
    -2\pi G \alpha \rho_1 \rho_2 S  \lambda^2 e^{-d/\lambda} \left(1 - e^{-t_1/\lambda}\right)\notag\\
&\times    
    \left(1 - e^{-t_2/\lambda}\right).
\end{align}
Edge effects, present for finite plates with half-lengths $x$ and $y$, introduce corrections    
\begin{align}
    F_{\rm Yuk,pp}(d) = F_{\rm Yuk-i,pp}(d)   \chi_{\rm edge}(x/2)\chi_{\rm edge}(y/2)~,
\end{align}
where $\chi_{\rm edge}(L)=1-(\lambda/L)\left(1-e^{-L/\lambda}\right)$ is the correction factor.
This factor ensures that the Yukawa force correctly scales with the volume for large $\lambda \gg x,y,t_1, t_2$. In the limit $\lambda \ll x,y$, $\chi_{\rm edge} \to 1$ and the infinite-plate result is recovered. 
The observable force gradient $F'_{\rm Yuk}(d)$ is found by differentiating $F_{\rm Yuk}(d)$.

For the s-s Casimir experiment geometry with spheres of radii $R_1$ and $R_2$ an exact calculation is more complex. However, for separations and interaction ranges much smaller than the sphere radii  $d, \lambda \ll R_1, R_2$, the PFA modeling s-s as effectively  p-p configurations with corrections provides an excellent estimate, as shown for the s-p configuration~\cite{Decca:2009fg}. The PFA relates the force between two curved surfaces to the interaction energy per unit area  $W(d)$  between two parallel plates  via the Derjaguin approximation 
\begin{align}
    F^{\text{PFA}}_{\rm ss}(d) = 2\pi R_{\rm eff} W_{\rm pp}(d),
\end{align}
where $R_{\rm eff} = (R_1^{-1} + R_2^{-1})^{-1}$ is the effective radius. 

For a Yukawa potential, the interaction energy per area between two semi-infinite plates is
\begin{align}
    W(d)_{\text{Yuk,pp}} = -2\pi G\alpha \rho_A\rho_B\lambda^3 \exp(-d/\lambda).
\end{align}
Thus, the PFA s-s Yukawa force gradient is
\begin{align} 
    F'^{\rm ~PFA}_{\rm Yuk,ss}(d) = 4\pi^2 G \alpha \lambda^2 R_{\rm eff} \rho_A \rho_B \exp(-d/\lambda). \label{eq:PFA_0}
\end{align}

Eq.~\eqref{eq:PFA_0}  makes explicit that the Yukawa signal scales differently in the s-s and p-p, $F'^{~\rm PFA}_{\rm Yuk,ss}(d) \propto  \lambda^2 e^{-d/\lambda}$  versus $F'_{\rm Yuk,pp}(d) \propto \lambda e^{-d/\lambda}$. This geometry dependence is central to the present work, implying that different Casimir configurations probe distinct functional dependencies of hypothetical interactions relative to the Casimir background. As a result, geometry provides additional means to disentangle new force contributions beyond capabilities of a single experimental configuration. Hence, different geometries have complementary sensitivities to different interaction ranges $\lambda$, with the p-p setup being more sensitive at shorter ranges, while the s-s configuration gains sensitivity at longer ranges due to the $\lambda^2$ scaling. This enables distinguishing new force contributions that affect such scalings.
While illustrative, simplified PFA does not account for the possible complex structure of the experimental configuration and its validity is limited to the case $d,~\lambda\ll R_0$ with $R_0$ being the characteristic length scale of the object.

Unlike the Casimir force, which is primarily a surface effect for conductive materials,
a hypothetical Yukawa-type interaction is volumetric and depends on the full mass distribution
of the interacting objects. To model realistic multilayer configurations, we compute Yukawa contributions using density superposition techniques. We numerically verify our results and accurate modeling of geometry, as discussed in Supplemental Material.

\textit{Casimir force experiments.--}
\label{sec:exp_config}
Precision Casimir experiments probe forces at microscopic separations of objects by exploiting quantum vacuum fluctuations of the electromagnetic field between their surfaces. The Casimir pressure $P_{\rm C}$ at finite temperature can be calculated using Lifshitz theory~\cite{Lifshitz:1956zz,Dzyaloshinskii:1961sfr}, see Supplemental Material.

For our analysis, Casimir forces constitute the dominant effective background, which we compute
in detail together with the Yukawa-type new force contributions in a multi-layered material as described in the Supplemental 
Material.
We then apply this to measurements of two complementary experiments. The p-p experiment of Bressi 
et al.~\cite{Bressi:2002fr} measured forces between parallel plates in high 
vacuum, using a rectangular silicon cantilever 
($1.9~\text{cm} \times 1.2~ \text{mm} \times 47~ \mu\text{m}$) as resonator 
opposite a rigid silicon plate ($t_2 = 0.5~ \text{mm}$), both coated with 
$50~ \text{nm}$ Cr. Plate separation was electrostatically calibrated to 
$\pm 35~ \text{nm}$. Since the Yukawa interaction is volumetric, we model the 
plates as homogeneous silicon ($\rho_\text{Si} \simeq 2.33 \text{~g~cm}^{-3}$), 
with the thin Cr coating contributing negligibly to total mass.

The s-s measurement of Garrett et al.~\cite{Garrett:2018zpo} operated in 
ambient air, with hydrodynamic damping subtracted via lock-in detection 
exploiting the $90^\circ$ phase difference between conservative and dissipative 
forces. We focus on the sphere pair with radii $R_1 = 34.2~\mu\text{m}$ and  
$R_2 = 46.9~\mu\text{m}$, which provide the most stringent constraints. The 
spheres are hollow glass microspheres (Trelleborg SI-100) coated with 
$\text{Cr}(3 ~\text{nm})/\text{SiO}_2(50 ~\text{nm})/\text{Cr}(3 ~\text{nm})/%
\text{Au}(100~ \text{nm})$. For the volumetric Yukawa calculation we model 
each sphere as three concentric shells consisting of an outer Au layer 
($t_\text{Au} = 100~\text{nm}$, $\rho_\text{Au} = 19.3~\text{g~cm}^{-3}$), 
an intermediate $\text{SiO}_2$ shell 
($t_{\text{SiO}_2} \simeq 546 ~\text{nm}$, 
$\rho_{\text{SiO}_2} \simeq 2.6 ~\text{g~cm}^{-3}$), and a hollow vacuum core. 
The dense Au coating dominates the Casimir interaction while the low-density 
core critically reduces the effective mass for Yukawa sensitivity.

\begin{figure*}[t]
    \centering
    \includegraphics[width=0.48\linewidth]{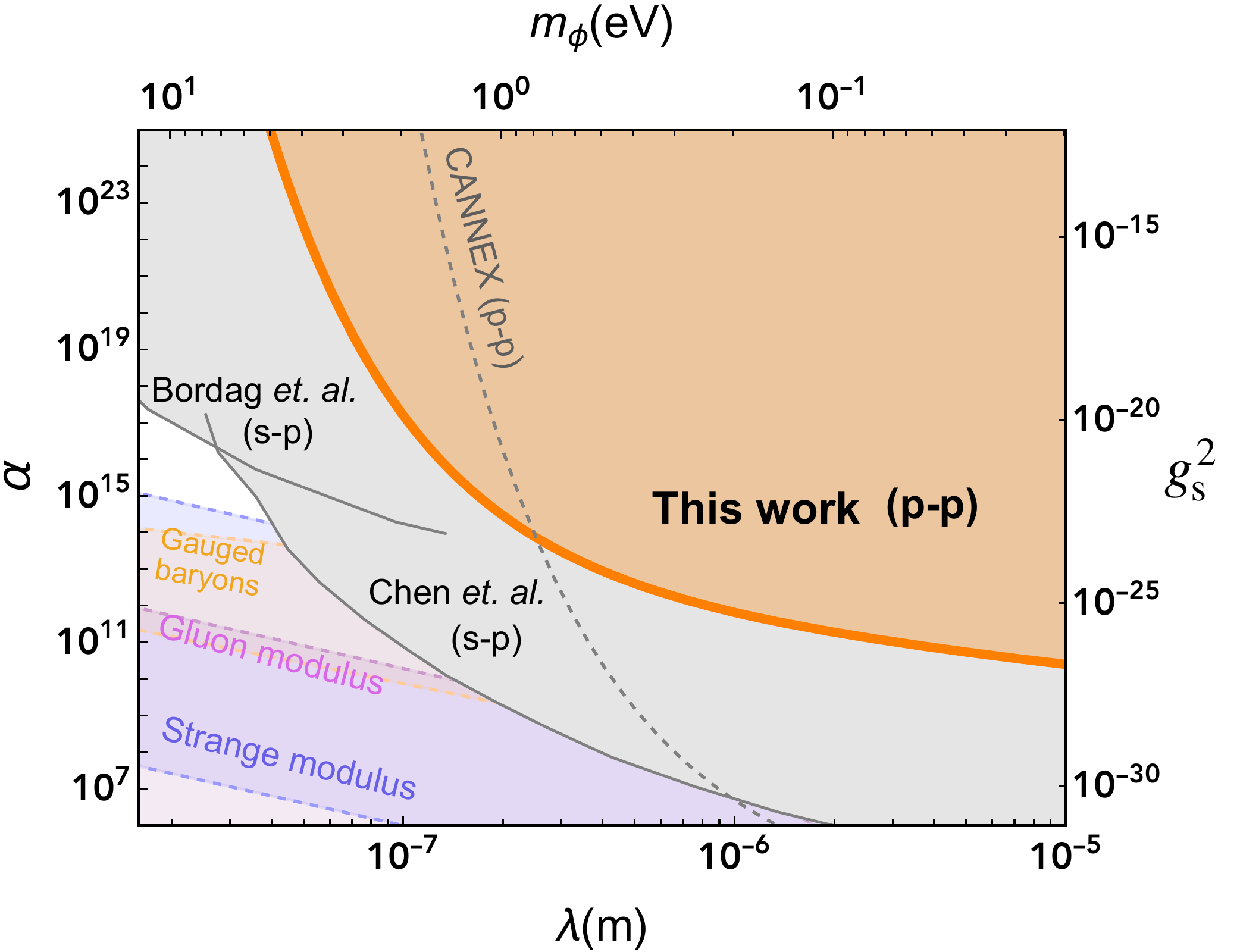}
    \includegraphics[width=0.48\linewidth]{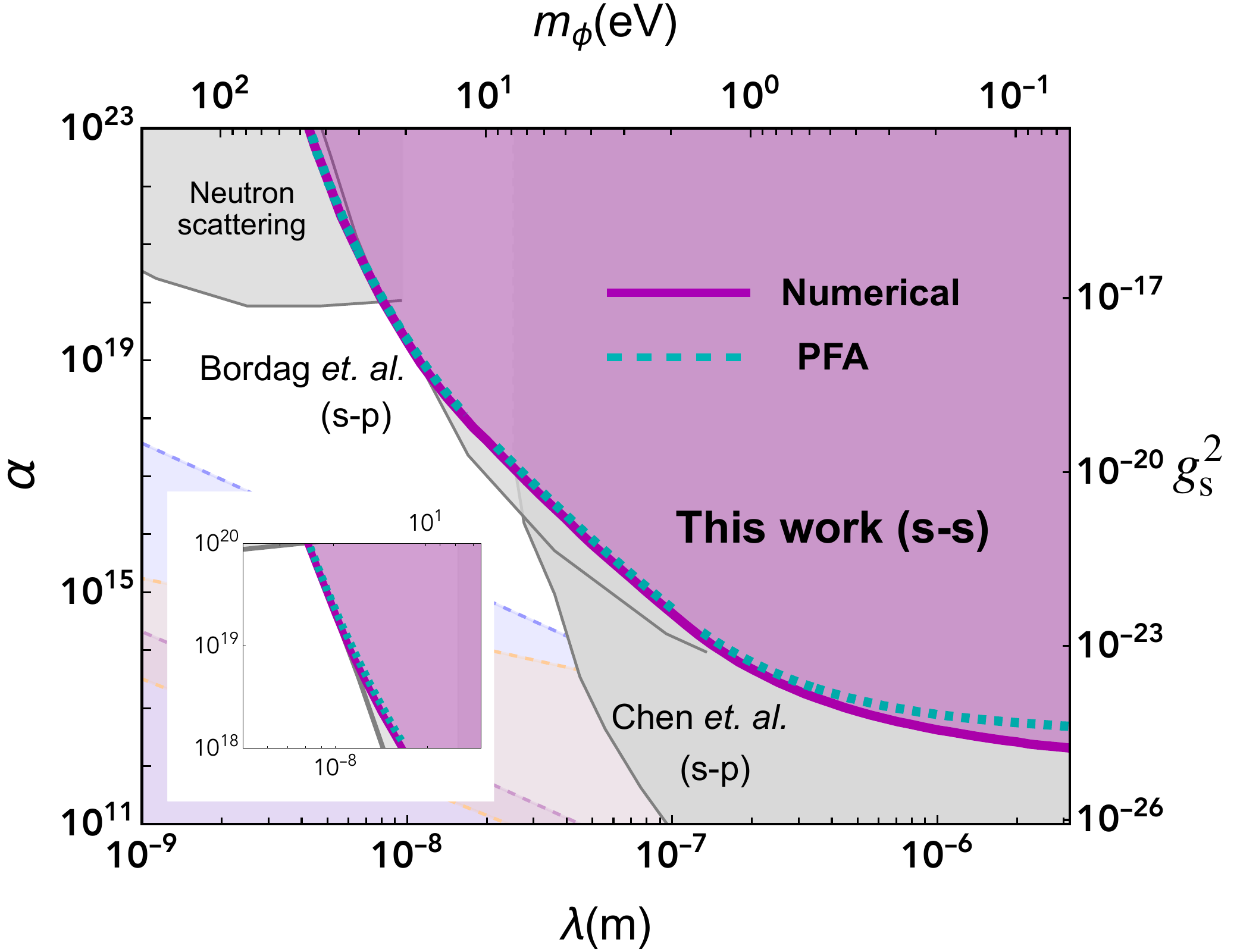}
    \caption{\textbf{[Left]} Constraints on the Yukawa coupling strength $\alpha$ as a function of characteristic force distance from the p-p Casimir geometry. Previous limits from s-p Casimir experiments ~\cite{Chen:2014oda, Bordag:2001qi}, as well as sensitivity projections for the proposed p-p CANNEX experiment~\cite{Sedmik:2020cfj} are shown. The right axis corresponds to coupling strength $g_s$ of the scalar mediator with protons and neutrons assuming universal coupling, while the top axis denote the corresponding mediator mass.  \textbf{[Right]} New constraints on the Yukawa coupling strength $\alpha$ from the s-s Casimir geometry.  The solid line indicates our numerical calculations of the Yukawa-type force, while the dotted line shows the  PFA results we obtained considering multilayered spheres. Example representative benchmarks from simplified models with gauged baryons, gluon-type modulus and strange modulus where effective $\alpha \gg 1$ can be predicted are displayed.}
    \label{fig:limits}
\end{figure*}

\textit{Plate-plate measurements.--}  
In the p-p in vacuum experiment of Bressi et al.~\cite{Bressi:2002fr}, the squared cantilever oscillation frequency shift
$\Delta\nu^2(d) = \nu^2 - \nu_0^2$ with oscillations detected by interferometer for  free frequency $\nu_0$ and separation distance $d$ can be modeled as the sum of contributions from the Casimir force and a residual electrostatic force
\begin{equation}
    \Delta\nu^2(d) = -\frac{1}{4\pi^2 m_{\text{eff}}}
    \left(
         F_{\text{C}}'
        +
        F_{\text{el}}'
    \right) =
    -\frac{C_{\rm el}V_r^2}{d^3}
    -\frac{C_{\rm C}}{d^5}~,
    \label{eq:freq_shift_pp}
\end{equation}
where $m_{\text{eff}}$ is the effective mass of the cantilever mode and $V_r$ is the residual voltage between the plates. The coefficients
$C_{\rm el}=\epsilon_0 S/(4\pi^2 m_{\text{eff}})$ and $C_{\rm C}$ characterize the electrostatic and Casimir contributions, respectively, with $S$ denoting the effective interaction area and $\epsilon_0$ is the vacuum dielectric constant.
The electrostatic coefficient
$C_{\rm el}=(4.24\pm0.11)\times10^{-13} ~{\rm Hz}^2{\rm ~m}^3{\rm ~V}^{-2}$
is obtained by varying large applied voltages and fitting the dependence of the frequency shift on $V_r$. This procedure provides an electrostatic calibration of the cantilever mode,
$1/(4\pi^2 m_{\rm eff})=C_{\rm el}/(\epsilon_0 S)$.
After subtracting the electrostatic contribution, the Casimir coefficient
$C_{\rm C}=(2.34\pm0.34)\times10^{-28} {\rm ~Hz}^2{\rm ~m}^5$ is extracted from the distance dependence of the residual frequency shifts and agrees with the theoretical prediction at the $\sim15\%$ level. 
 
A dominant systematic uncertainty arises from the uncertainty in the plate separation $\sigma_d$. Since the full covariance matrix of the dataset is not available, we adopt a conservative treatment using the effective variance method~\cite{Orear:1981qt}. Assuming that the distance uncertainty and frequency measurement uncertainty are independent, the effective uncertainty in the force gradient induced frequency shift $\sigma_{\text{eff}}$ is given by
 \begin{equation} \label{eq:sigmaeff}
    \sigma_{\text{eff}}^2
    =
    \sigma_{\Delta \nu_C^2}^2 + \sigma_{d}^2\big| (\Delta\nu_C^2)'\big|^2_{d=1.1~\mu\text{m}} \simeq (23.44~{\rm Hz}^2)^2 ,
\end{equation}
where $\sigma_{\Delta \nu_C^2}^2$ is the reported uncertainty of square frequency shift while $\sigma_{d}$ is the  separation distance uncertainty. Subtracting the electrostatic contributions as in Ref.~\cite{Bressi:2002fr}, we obtain $(\Delta \nu_C^2)'$ by differentiating the fitted $\Delta\nu_C^2(d) = -1/(4\pi^2 m_{\text{eff}})
         F_{\text{C}}' = -C_C/d^5$ that denotes the  
total dominant effective residue background of frequency shift for Yukawa-type signal. By varying $\sigma_{\rm eff}(d)$ with $d$, we find that measurements at $d \simeq 1.1 ~\mu$m are the most robust providing minimum total uncertainty. We hence use data at $d = 1.1~\mu$m for probing new force effects.

A hypothetical Yukawa-type force would induce an additional frequency shift
\begin{equation}
    \Delta\nu^2_{\text{Yuk}}(d)
    =
    -\frac{1}{4\pi^2 m_{\text{eff}}}
    F'_{\rm Yuk,pp}(d).
    \label{eq:freq_shift_yuk_pp}
\end{equation}
Constraints on the coupling strength $\alpha$ are obtained by requiring this contribution to remain within the experimental uncertainty
\begin{equation}
    \left|
        F'_{\rm Yuk,pp}(d)
    \right|
    \le
    \sigma_{\text{eff}}(d).
    \label{eq:pp_constraint}
\end{equation}
For each interaction range $\lambda$, we compute the corresponding upper bound on $\alpha$. We have verified (see Supplemental Material) that Casimir edge effects only contribute at sub-percent level, consistent with findings of Ref.~\cite{Gies:2006xe}. 

In Fig.~\ref{fig:limits} we display the resulting p-p limits on hypothetical Yukawa-type forces as a function of the interaction range $\lambda$. The final exclusion curve is obtained as the envelope of the most stringent limits across all measured separations.  Our analysis provides the first constraints derived from the p-p Casimir geometry, as originally proposed for Casimir effects. While current p-p bounds are not yet as constraining as leading s-p measurements, they establish the parametric sensitivity of this geometry and define a new experimental direction with distinct scaling behavior, directly relevant for next generation parallel plate searches such as CANNEX~\cite{Sedmik:2020cfj}.

\textit{Sphere-sphere measurements.--}  
The s-s Garrett et al. experiment~\cite{PhysRevLett.120.040401} measures the force gradient between two spheres and normalizes it by the effective radius
$R_{\rm eff}$.
The measurement is compared to the Casimir pressure between parallel plates, $P_{\rm C,pp}$, through the parameterization
\begin{equation}
    \frac{1}{R_{\rm eff}} F'
    =
    2\pi P_{\rm C,pp}
    \left(
        1
        +
        \frac{\beta' d}{R_{\rm eff}}
        + \dots
    \right),
    \label{eq:beta_prime_def}
\end{equation}
where $\beta'$ quantifies deviations from PFA.
No statistically significant deviation from the PFA prediction is observed, constraining from combined measurements  
$\beta'=-6\pm27$ at the $2\sigma$ level, with $1\sigma$ uncertainty
of $\sigma_{\beta'}=13.5$. While experiment is done in air, hydrodynamic damping contributions are subtracted. Our calculation of Casimir force (see Supplemental Material) reproduces Ref.~\cite{Garrett:2018zpo}.

A new Yukawa-type interaction would contribute an additional force gradient $F'_{\rm Yuk,ss}$, beyond PFA. Requiring this to remain within the allowed deviation  yields
\begin{equation}
    \left|
        F'_{\rm Yuk,ss}
    \right|
    \le
    2\pi R_{\rm eff} P_{\rm C,pp}
    \frac{\sigma_{\beta'} d}{R_{\rm eff}}.
    \label{eq:ss_constraint_raw}
\end{equation}
Using PFA of Eq.~\eqref{eq:PFA_0} this reduces to
\begin{equation}
    \alpha
    \le
    \frac{
        \sigma_{\beta'} d P_{\rm C,pp}(d)
    }{
        2\pi G \lambda^2 \rho_A \rho_B
    }
    e^{d/\lambda}.
    \label{eq:alpha_limit_ss}
\end{equation} 
Since PFA corrections scale differently from Yukawa-type force contributions, they are distinguishable.

Specifically, we calculate the interactions using the pair of spheres with radii $R_1 = 34.2~\mu\text{m}$ and $R_2 = 46.9~\mu\text{m}$, giving   $R_{\rm eff} = 19.78~\mu\text{m}$, corresponding to the measurements taken at various separations shown in Ref.~\cite{Garrett:2018zpo}. We calculate the Casimir pressure $F_{\rm C,pp}$ between Au-coated surfaces at ambient temperature using Lifshitz theory~\cite{Lifshitz:1956zz}, as discussed in the Supplemental Material. Using the empirical fit from the experiment, $P_{\rm C,pp}(d)\propto d^{-3.7}$, the sensitivity scales approximately as $\alpha\propto d^{-2.7}e^{d/\lambda}$. For a given $\lambda$, this expression is minimized near $d\simeq2.7\lambda$, which corresponds to the separation of maximal sensitivity. We therefore evaluate Eq.~(\ref{eq:alpha_limit_ss}) at the optimal distance within the experimental range $d\in[0.05, 0.3]~\mu\text{m}$, adopting the most stringent bound as our resulting limit.

In Fig.~\ref{fig:limits} we show the resulting  s-s constraints on Yukawa-type interactions. The s-s geometry yields the strongest constraints within the Casimir experiments at interaction ranges $\lambda \lesssim 10^{-8}~\mathrm{m}$, demonstrating that alternative geometries can outperform the conventional configuration in specific regimes. Our calculations of Yukawa-type force contributions account for the multilayer structure of the spheres, including the finite thickness of the gold and silica layers, as detailed in the Supplemental Material. 
We also display comparison of numerical results with PFA.

At very short interaction ranges ($\lambda < 100~\mathrm{nm}$), the interaction is dominated by the dense gold shells ($\rho_{\mathrm{Au}} \simeq 19.3~\mathrm{g~cm^{-3}}$). As $\lambda$ increases, the lower-density glass core reduces the effective coupling relative to that expected for solid gold spheres. The resulting limits are particularly competitive in the $10 -100~\mathrm{nm}$ range. Since the Yukawa-type signal depends sensitively on deviations from PFA, improved experimental control of  uncertainties could significantly enhance sensitivity in the s-s geometry. At very short distances the sensitivity is limited by the minimum experimental separation.

A broad variety of motivated extensions of the Standard Model predict Yukawa-type interactions at submicron scales, including models with extra dimensions involving radions or dilatons, as well as scenarios with composite or extended gravitons~\cite{Adelberger:2003zx,Antoniadis:1998ig,Sundrum:2003jq,Murata:2014nra}. In theories involving moduli or scalar mediators, the effective macroscopic coupling relative to gravity can be significantly enhanced  allowing $\alpha \gg 1$. The benchmark regions shown in Fig.~\ref{fig:limits} correspond to simplified scenario examples. In these scenarios, $\alpha$ should be interpreted as an effective macroscopic coupling normalized to gravity, rather than a fundamental microscopic parameter. Since gravity is extremely weak, even modest underlying particle couplings can correspond to very large effective values of $\alpha$, making this regime a natural target for precision laboratory searches. Our results are also relevant for ``screened'' forces, such as in chameleon and symmetron models, in which nonlinear screening dynamics suppress the force in dense environments while allowing larger effective couplings in laboratory conditions~\cite{Khoury:2003rn,Hinterbichler:2010es,Burrage:2021nys}.

\textit{Conclusions.--}
\label{sec:conclusion}
We have shown that Casimir measurements performed in alternative geometries provide a novel probe of short range interactions parameterized as deviations from Newtonian gravity. While previous Casimir-based searches have almost exclusively relied on the s-p configuration, s-s and p-p geometries exhibit distinct geometric scalings of Yukawa-type signals relative to the Casimir background, geometry enables additional separation of bulk new interactions from surface-dominated quantum effects.

By performing volumetric calculations that incorporate realistic multilayer mass distributions, we reanalyzed precision measurements in different geometries and derived the first constraints from s-s and p-p Casimir setups. Our results complete the canonical set (s-s, s-p, p-p) of Casimir geometries for new force searches. Together, this defines a unified framework in which geometry acts as an additional observable, allowing additional systematic disentangling of new physics contributions across experimental setups.
 In particular, the s-s configuration yields the strongest Casimir-based bounds for $\lambda \lesssim 10^{-8}~\mathrm{m}$, demonstrating that alternative geometries can outperform the conventional configuration. More broadly, these results demonstrate that Casimir geometry itself provides a key independent  ingredient, rather than secondary experimental detail, which can be leveraged for isolating novel weak interactions from the dominant quantum vacuum  background.

\textit{Acknowledgments.--} 
This work is supported by the World Premier International Research Center Initiative (WPI), MEXT, Japan. 
\bibliography{references}
\phantom{i}

\appendix

\clearpage
\onecolumngrid

\centerline{\large {Supplemental Material for}}
\medskip
{\centerline{\large \bf{Casimir Geometry as a Probe of Short Range Forces}}} 
\medskip
{\centerline{Xiaolin Ma, Volodymyr Takhistov, Hideo Iizuka}}
\bigskip
\bigskip

In this Supplemental Material we provide additional details on the calculation of Casimir forces for the p-p and s-s geometries.

\subsection*{A.~Frequency Shifts in Oscillating Cantilever  Experiments}

Oscillating cantilevers are often employed in experimental measurements of Casimir force effects, typically exerted between a cantilever and a material surface. Such configurations include p-p measurements of Bressi et al.~\cite{Bressi:2002fr} and s-s measurements of Garrett et al.~\cite{Garrett:2018zpo}, where in the latter case one of the spheres is attached to an atomic force microscope cantilever.
The cantilever can be modeled as a harmonic oscillator with effective mass $m_{\rm eff}$ and intrinsic spring constant $k_0$. In the absence of external interactions, small displacements $x$ about equilibrium satisfy  
\begin{equation}
    m_{\text{eff}} \ddot{x} + k_0 x = 0
\end{equation}
with the intrinsic resonant frequency $\nu_0$ given by
\begin{equation}
 \nu_0^2 = \dfrac{k_0}{4\pi^2 m_{\rm eff}}~.
 \end{equation}

 In the presence of a distance-dependent external force $F_{\rm ext}(z)$, where $z$ is the plate separation in the case of the p-p setup, we consider small amplitude oscillations about a  separation distance $d$, such that $z = d + x$ with $|x| \ll d$. Taylor expanding the force to linear order 
\begin{equation}
    F_{\rm ext}(d+x) \simeq F_{\rm ext}(d) + x F_{\rm ext}'(d) + \mathcal{O}(x^2) 
\end{equation}

The constant term $F_{\rm ext}(d)$ shifts the static equilibrium position. Redefining $x \to x + F_{\rm ext}(d)/k_0$ shows that this does not affect the oscillation frequency. The linear term modifies the effective restoring force, yielding the equation of motion
\begin{equation}
    m_{\rm eff}\ddot{x} + \left(k_0 -    F_{\rm ext}'(d)\right)x = 0   .
\end{equation}
The modified resonant frequency is therefore
\begin{equation}
\nu^2 = \frac{1}{4\pi^2 m_{\rm eff}} \left(k_0 - F_{\rm ext}'(d)\right)~.
\end{equation}
Hence, to leading order in the force gradient, the measured shift in the squared frequency is
\begin{equation}
    \Delta \nu^2 = \nu^2 - \nu_0^2 = -\frac{1}{4\pi^2 m_{\rm eff}}  F_{\rm ext}'(d)  .
\end{equation}

\subsection*{B.~Casimir Force Between Two Spheres}
\label{sec:casimir_calc}

To determine sensitivity to Yukawa interactions, the electromagnetic Casimir force background must be appropriately modeled.
The Casimir pressure between parallel plates separated by a distance $d$ at finite temperature can be obtained from Lifshitz theory~\cite{Lifshitz:1956zz,Klimchitskaya:2009cw,Iizuka:2019ovv,Harris:2000zz,Klimchitskaya:2009dw} as
\begin{equation}  
    P_{\rm C, pp} =  -\frac{T}{\pi} \sum_{n=0}^{\infty}{'} \int_{0}^{\infty} k_{\parallel} dk_{\parallel} k_{z,n} \Bigg[ \left( \frac{e^{2k_{z,n} d}}{\mathcal{R}_{\text{TM}}^2(i\xi_n)} - 1 \right)^{-1}    + \left( \frac{e^{2k_{z,n} d}}{\mathcal{R}_{\text{TE}}^2(i\xi_n)} - 1 \right)^{-1} \Bigg], 
\label{eq:Lifshitz_FiniteT}\end{equation} 
where  $k_{\parallel}$ is the lateral wavevector component, and the prime on the summation indicates that the $n=0$ term is weighted by a factor of $1/2$.  
Here, we included for precise calculation the summation over discrete Matsubara frequencies $\xi_n  =  2\pi n  T$ that are relevant for finite temperature effects, defined for integers $n = 0, 1, 2 \dots$.
The variables $k_{z,n} = \sqrt{k_{\parallel}^2 + \xi_n^2}$ denote   the wavevector components in vacuum along the separation axis for the $n$-th mode, and the   reflection coefficients $\mathcal{R}_{\alpha}$ ($\alpha = \text{TM, TE}$) account for the layered structure of the materials.

For the s-s configuration, we independently compute the Casimir background using Lifshitz theory with realistic material properties and map the parallel plate result to the s-s geometry using PFA, following the experimental convention of Garrett et al.~\cite{PhysRevLett.120.040401}.
Ref.~\cite{PhysRevLett.120.040401} reports measurements in terms of a relative deviation parameter $\beta'$. We reproduce the Casimir pressure $P_{\rm C,pp}$ between parallel plates used in their data analysis  and   consistently translate the experimental bounds into constraints on an additional Yukawa-type interactions.

The experimental s-s configuration of Ref.~\cite{Garrett:2018zpo}  involves finite thickness gold coatings $t=100~{\rm nm}$  deposited on silicon substrates. Accordingly, the reflection coefficients entering the Lifshitz formula of Eq.~\eqref{eq:Lifshitz_FiniteT} are those of a layered system with vacuum, Au and Si. The   reflection coefficients $\mathcal{R}_\alpha$ are obtained using 
\begin{equation}
    \mathcal{R}_{\alpha}(i\xi_n) = \frac{r_{01,\alpha} + r_{12,\alpha} e^{-2 k_{1,z,n} t}}{1 + r_{01,\alpha} r_{12,\alpha} e^{-2 k_{1,z,n} t}},
\end{equation}
where $r_{ij,\alpha}$ are Fresnel coefficients at the interface between media $i$ and $j$ (0: vacuum, 1: Au, 2: Si), and
\begin{equation}
k_{1,z,n} = \sqrt{k_\parallel^2 + \epsilon_{\rm Au}(i\xi_n)\,\xi_n^2}  
\end{equation}
is the perpendicular wavevector component inside the gold layer.

The dielectric response functions $\epsilon(i\xi)$ are evaluated along the imaginary frequency axis. For gold, we adopt a Drude-Lorentz model that captures the behavior over wide range of frequencies
\begin{equation}
\epsilon_{\rm Au}(i\xi) =
1 + \frac{\omega_p^2}{\xi(\xi + \Gamma)}
+ \sum_{j=1}^{N} \frac{f_j \omega_p^2}{\omega_j^2 + \xi^2 + \gamma_j \xi} \, ,
\end{equation}
with plasma frequency $\omega_p = 8.84~{\rm eV}$ and damping rate $\Gamma = 42~{\rm meV}$, consistent with Ref.~\cite{PhysRevLett.120.040401}. The multi-oscillator Lorentz terms account for interband transitions relevant for higher frequencies, and we fit parameters to tabulated optical data of Ref.~\cite{palik1998handbook}. At low frequencies $\xi \lesssim 1$~eV the response is dominated by the Drude model contributions. We employ $N=5$, which provides a stable and causal representation of $\epsilon_{\rm Au}(i\xi)$ for the Matsubara summation.

For the silicon sphere substrate, we use a standard multi-oscillator representation
\begin{equation}
\epsilon_{\rm Si}(i\xi) =
1 + \sum_k \frac{C_k}{\omega_k^2 + \xi^2 + g_k \xi} \, ,
\end{equation}
with parameters taken from optical data fits of Ref.~\cite{palik1998handbook}. This treatment consistently incorporates the effect of the finite Au coating and the dielectric substrate.
  
\begin{figure}[t]
    \centering
    \includegraphics[width=0.45\linewidth]{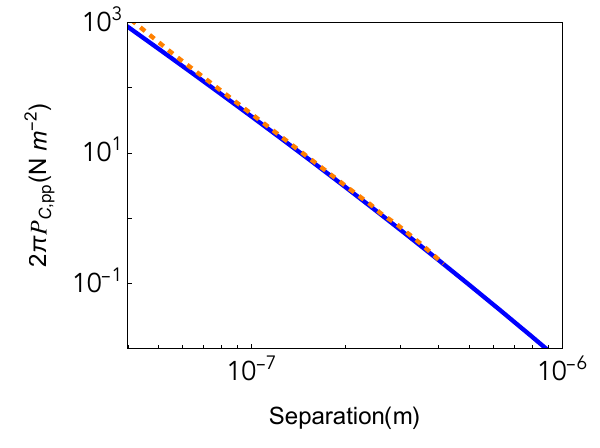}
    \caption{Casimir pressure of the s-s setup from our computation (solid blue line) compared with the theoretical result (orange dashed line) from Ref.~\cite{PhysRevLett.120.040401} [Fig.~3(c)], for the same experimental configuration.}
    \label{fig:placeholder}
\end{figure}

The resulting Casimir pressure is shown in Fig.~\ref{fig:placeholder}, where we reproduce to good agreement the theoretical curve of Ref.~\cite{PhysRevLett.120.040401}. The agreement validates our implementation and ensures consistency in limits on additional forces.
We note that there are corrections beyond PFA. Exact computations of the Casimir interaction between two spheres have been developed in Ref.~\cite{Bimonte:2018alu}. In our setup, the relevant expansion parameter is $a/R_{\rm eff} \ll 1$, where $a$ is the separation. The corresponding curvature corrections are therefore at the sub-percent level for metallic surfaces at room temperature, and remain well below both experimental uncertainties and theoretical modeling uncertainties. We thus adopt the PFA throughout.

\subsection*{C. Casimir Force Edge Effects in Plate-Plate Geometry}

Finite size effects can modify the Casimir pressure in realistic p-p configurations. For plates of finite extent $L \times L$, transverse translational invariance is broken and the momentum spectrum becomes discretized. In the continuum limit, the standard Lifshitz result is recovered, while finite size corrections arise from boundary contributions. The leading correction can be understood parametrically. The bulk Casimir pressure scales with the plate area, whereas edge contributions scale with the plate perimeter. This implies a relative correction of order
\begin{equation}
\frac{\Delta P_{\rm edge}}{P_{\rm C,pp}} \sim \mathcal{O}\!\left(\frac{d}{L}\right),
\end{equation}
where $d$ is the plate separation. Physically, this scaling reflects that the dominant transverse momenta contributing to the Lifshitz integral are $k_\parallel \lesssim 1/d$, due to exponential suppression at large momenta.

For the experimental configurations considered here  $d \ll L$, and the resulting edge corrections are at the sub-percent level. This is consistent with previous analysis of Casimir edge effects~\cite{Gies:2006xe}, and does not significantly affect  our main results.

\subsection*{D.~Analysis of Yukawa-like Forces in Multilayer Experiments}
\label{sec:numerical}

Realistic experimental configurations often involve multiple material layers. We describe a methodology for computing Yukawa-like force contributions accounting for multilayered setups.

\subsubsection{Superposition Method for Multilayered Spheres}

To efficiently evaluate effects of new Yukawa-like interactions in multilayered geometries, we employ a density contrast superposition method. 
For a s-s setup, as relevant for our analysis, a sphere with a radial density profile can be decomposed into a sum of concentric homogeneous spheres with effective density contrasts. For a structure with a core density $\rho_{\rm core}$ and a shell density $\rho_{\rm shell}$, this corresponds to a solid sphere of radius $R$ with density $\rho_{\rm shell}$, supplemented by a concentric sphere of radius $R_c = R - t_{\rm shell}$ with density contrast $\Delta \rho = \rho_{\rm core} - \rho_{\rm shell}$. 

This construction generalizes to more complex multilayer systems. For a configuration with layer boundaries $R_i$ and densities $\rho_i$ (ordered from outermost to innermost), the total force can be written as
\begin{equation}
    F_{\rm total} = F(R_1,\rho_1) + \sum_{i} F(R_i, \rho_i - \rho_{i-1})   .
\end{equation}
For a hollow core, the innermost density is set to zero, ensuring proper subtraction of the interior mass.

To calculate the Yukawa-like force gradient for the complex hollow, multilayered sphere geometry with different densities,  we decompose each physical sphere $k$ (where $k=A,B$) into a sum of three concentric, homogeneous solid spheres with effective densities $\Delta\rho_{k,n}$ and radii $R_{k,n}$. We consider the s-s experiment from Ref.~\cite{Garrett:2018zpo} and model the setup as a three-layer system, with each sphere consisting of an outer Au shell, an intermediate $\text{SiO}_2$ layer, and a hollow core. We therefore decompose each physical sphere $k = A,B$ into three effective components with radii $R_{k,i}$ and density contrasts
\begin{align}
    \Delta\rho_{k,1} &= \rho_{\rm Au}, \nonumber\\
    \Delta\rho_{k,2} &= \rho_{\rm SiO_2} - \rho_{\rm Au}, \nonumber\\
    \Delta\rho_{k,3} &= -\rho_{\rm SiO_2}   .
\end{align}

Besides the numerical calculations, this method also allows for an efficient calculation of the Yukawa force using PFA method, which is valid for interaction ranges and separations much smaller than the sphere radii ($\lambda, d \ll R_{1,2}$). The validity of the PFA for this geometry has been confirmed by comparing with direct numerical  integrations. 
Within PFA, the total Yukawa-like force gradient is obtained by summing pairwise contributions between every effective component $i$ of the top sphere and every component $j$ of the bottom sphere
\begin{equation}
    F'_{\rm total}(d) = \sum_{i,j} \Delta\rho_{A,i} \Delta\rho_{B,j} 
    \mathcal{G}  \left(d_{ij}, R_{{\rm eff},ij}\right)   ,
\end{equation}
where $R_{{\rm eff},ij} = (R_{A,i}^{-1} + R_{B,j}^{-1})^{-1}$ and $d_{ij} = d + \delta_i + \delta_j$ is the surface to surface separation between the effective sub-spheres. The function $\mathcal{G}$ denotes the Yukawa-like force gradient for homogeneous spheres, see Eq.~\eqref{eq:PFA_0}.
This analytical method allows for rapid computations of the parameter space $(\alpha, \lambda)$, and gives results consistent with high accuracy with more computationally intensive numerical methods in the appropriate PFA-valid limits. Here, the impact of uncertainties in the sphere radii is subleading. For a relative radius uncertainty $x = \delta R / R$, the PFA scaling implies a fractional change in sensitivity of order $x R_2/(R_1+R_2)$, which is further suppressed in the final results after error propagation.

\subsubsection{Comparison of Numerical Computation and PFA}
To verify the PFA results, particularly at larger $\lambda$ where surface curvature effects become significant, we perform a direct numerical integration of the Yukawa-type interaction. The mass-normalized Yukawa-type force and its radial derivative between two point masses  separated by distance $r$ are given by 
\begin{align}
    F_{\rm Yuk}(r) &= G \alpha  \frac{e^{-r/\lambda}}{r} \left( \frac{1}{r} + \frac{1}{\lambda} \right)\hat{\bm r},\nonumber\\
    F'_{\rm Yuk}(r) &= -G \alpha  e^{-r/\lambda} \left( \frac{r^2+2\lambda r+2\lambda^2}{r^3\lambda^2}\right)\hat{\bm r}
    \label{eq:point_force}
\end{align}
Since the direction of the force and its derivative are along the line between two mass points, these quantities need to be projected along the $z$-axis connecting the sphere centers. 

The total Yukawa-type force gradient is obtained by integrating over the volumes $V_1$ and $V_2$
\begin{equation}
    F_{\rm Yuk, tot}' = \int_{V_1} d^3\bm{r}_1 \int_{V_2} d^3\bm{r}_2 F'_{\rm Yuk}(|\bm{r}_1 - \bm{r}_2|)   \rho(\bm{r}_1)\rho(\bm{r}_2) \hat{\bm{z}}   .
\end{equation}
We evaluate this integral using Monte Carlo sampling with importance weighting tailored to the exponential Yukawa-type kernel, $P(x)\propto e^{-x/\lambda}$ along the axial direction, while sampling transverse coordinates uniformly. The integration includes the appropriate Jacobian factors and shell geometries. 

We find that numerical results, based on considering $\mathcal{O}(10^6)$ samples, show excellent agreement with the PFA calculation in the regime $\lambda, d \ll R$. This confirms the validity of our analytical treatment. At larger $\lambda$, finite-size effects can lead to deviations from the PFA scaling, as expected. 

\end{document}